\documentclass{emulateapj}

\usepackage{color}
\usepackage{longtable}
\usepackage{color}
\usepackage{float}
\usepackage{natbib}
\usepackage{url}
\usepackage{hyperref}
\usepackage{breakurl}
\usepackage{perpage} %the perpage package
\MakePerPage{footnote} %the perpage package command
\citeindextrue

\shorttitle{WFC3 Grism Redshifts in GOODS-S}
\shortauthors{Morris et al.}

\begin{document}
 
 \title{A WFC3 Grism Emission Line Redshift Catalog in the GOODS-South Field}
 
 \author{Aaron M. Morris\altaffilmark{1}, 
 Dale D.~Kocevski\altaffilmark{1},
 Jonathan R.~Trump\altaffilmark{2,3},
 Benjamin J.~Weiner\altaffilmark{4},
 Nimish P.~Hathi\altaffilmark{5},
 Guillermo Barro\altaffilmark{6},
 Tomas Dahlen\altaffilmark{7}, 
 Sandra M.~Faber\altaffilmark{6},
 Steven L.~Finkelstein\altaffilmark{8},
 Adriano Fontana\altaffilmark{9}, 
 Henry C.~Ferguson\altaffilmark{7},
 Norman A.~Grogin\altaffilmark{7},
 Ruth Gr{\"u}tzbauch\altaffilmark{10,11}, 
 Yicheng Guo\altaffilmark{6,12},
 Li-Ting Hsu\altaffilmark{13},
 Anton M.~Koekemoer\altaffilmark{7},
 David C. Koo\altaffilmark{6},
 Bahram Mobasher\altaffilmark{14},
 Janine Pforr\altaffilmark{15,16},
 Mara Salvato\altaffilmark{13}, 
 Tommy Wiklind\altaffilmark{7},
 Stijn Wuyts\altaffilmark{13}}
 
 \altaffiltext{1}{Department of Physics and Astronomy, University of Kentucky, Lexington, KY 40506}
\altaffiltext{2}{Department of Astronomy and Astrophysics, Pennsylvania State University, University Park, PA 16802}
\altaffiltext{3}{Hubble Fellow}
\altaffiltext{4}{Steward Observatory, 933 N.~Cherry Street, University of Arizona, Tucson, AZ 85721, USA}
\altaffiltext{5}{Aix Marseille Universit\'{e}, CNRS, LAM (Laboratoire d'Astrophysique de Marseille) UMR 7326, 13388, Marseille, France}
\altaffiltext{6}{UCO/Lick Observatory and Department of Astronomy and Astrophysics, University of California, Santa Cruz, CA 95064 USA}
\altaffiltext{7}{Space Telescope Science Institute, Baltimore, MD 21218, USA}
\altaffiltext{8}{The University of Texas at Austin, 2515 Speedway, Stop C1400, Austin, Texas 78712, USA}
\altaffiltext{9}{INAF Osservatorio Astronomico di Roma, Via Frascati 33,00040 Monteporzio (RM), Italy}
\altaffiltext{10}{School of Physics \& Astronomy, University of Nottingham, Nottingham NG7 2RD}
\altaffiltext{11}{Centre for Astronomy and Astrophysics, University of Lisbon, P-1349-018 Lisbon, Portugal}
\altaffiltext{12}{Department of Astronomy, University of Massachusetts, Amherst, MA 01003, USA}
\altaffiltext{13}{Max-Planck-Institut f{\"u}r extraterrestrische Physik, Giessen-bachstrasse D-85748 Garching, Germany}
\altaffiltext{14}{Department of Physics and Astronomy, Colby College, Waterville, ME, USA}
\altaffiltext{15}{Institute of Cosmology and Gravitation, University of Portsmouth, Dennis Sciama Building, Burnaby Road, Portsmouth PO1 3FX, UK}
\altaffiltext{16}{NOAO, 950 N.~Cherry Avenue, Tucson, AZ 85719, USA}

\email{aaron.morris2@uky.edu}

\begin{abstract}

We combine \emph{HST}/WFC3 imaging and G141 grism observations from the CANDELS and 3D-HST surveys to produce a catalog of grism spectroscopic redshifts for galaxies in the CANDELS/GOODS-South field.  The WFC3/G141 grism spectra cover a wavelength range of $1.1\leq \lambda \leq 1.7 \mu m$ with a resolving power of $R\sim 130$ for point sources, thus providing rest-frame optical spectra for galaxies out to $z\sim3.5$.  The catalog is selected in the \emph{H}-band (F160W) and includes both galaxies with and without previously published spectroscopic redshifts.   Grism spectra are extracted for all \emph{H}-band detected galaxies with H$\leq$24 and a CANDELS photometric redshift $z_{phot}\geq 0.6$.  The resulting spectra are visually inspected to identify emission lines and redshifts are determined using cross-correlation with empirical spectral templates.  To establish the accuracy of our redshifts, we compare our results against high-quality spectroscopic redshifts from the literature.  Using a sample of 411 control galaxies, this analysis yields a precision of $\sigma_{NMAD}=0.0028$ for the grism-derived redshifts, which is consistent with the accuracy reported by the 3D-HST team.  Our final catalog covers an area of 153 arcmin$^{2}$ and  contains 1019 redshifts for galaxies in GOODS-S.  Roughly 60\% (608/1019) of these redshifts are for galaxies with no previously published spectroscopic redshift.  These new redshifts span a range of $0.677\leq z\leq 3.456$ and have a median redshift of $z=1.282$.  The catalog contains a total of 234 new redshifts for galaxies at $z>1.5$.  In addition, we present 20 galaxy pair candidates identified for the first time using the grism redshifts in our catalog, including four new galaxy pairs at $z\sim2$, nearly doubling the number of such pairs previously identified.

\end{abstract}

\keywords{catalogs, galaxies: high-redshift, techniques:spectroscopic}

\section{Introduction}

Accurate galaxy redshifts are vital to studying how the physical properties and environments of galaxies evolve over cosmic time. While high resolution, ground-based spectroscopy has traditionally provided the most reliable redshifts, these observations are time consuming for faint sources and are subject to the limited wavelength sensitivity of optical spectrographs, making it difficult to extent large redshift surveys beyond $z\sim1.2$ \citep{2007ApJ...660L...1D,2007ApJS..172...70L}.  Near-infrared (NIR) spectrographs are now pushing the redshift frontier into the so-called redshift desert \citep{2013ApJ...763L...6T,2014arXiv1408.2521S,2014arXiv1409.0447S,2014arXiv1409.6791W}, however at these wavelengths ground based observations are subject to contamination from atmospheric OH lines.  Photometric redshift estimates, on the other hand, can provide redshifts for large samples of galaxies, including relatively faint systems, at a lower observational cost compared to spectroscopy.  However, even the best photometric redshifts have errors of a few percent and are subject to catastrophic outliers for rare sources, such as active galactic nuclei (AGN), if their unique spectral energy distributions (SED) and/or strong emission lines are not properly modeled a priori \citep{2009ApJ...690.1250S,2011ApJ...742...61S}.

\begin{figure}
\epsscale{1.10}
\plotone{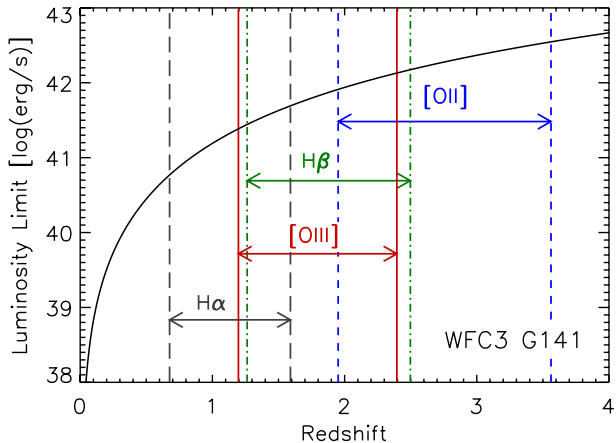}
\caption{Line luminosity limits for detecting the H$\alpha$, H$\beta$, [OIII], and [OII] emission lines with 2 orbits of the \emph{HST}/WFC3 G141 grism (corresponding to a line flux limit of $3\times10^{-17} {\rm erg/s/cm^{2}}$).  Dashed vertical lines indicate the redshift ranges at which various lines are visible in the G141 sensitivity window (1.1-1.7 $\mu$m).
\label{fig-redshift-range}}
\end{figure}

Slitless grism spectroscopy with the Wide Field Camera 3 (WFC3) onboard the Hubble Space Telescope (\emph{HST}) now provides a powerful alternative to ground-based spectroscopy and SED modeling for measuring distant redshifts.  The slitless nature of the WFC3/IR grism offers the ability to obtain a spectrum of each galaxy in the detector's field-of-view, while the significantly reduced background levels compared to the ground means emission lines can be detected for relatively faint galaxies with modest exposure times \citep[e.g.][]{2011ApJ...743..144T,2012ApJS..200...13B,2011AJ....141...14S,2011ApJ...743..121A,2011ApJ...742..111V}.  In addition, the near-infrared sensitivity of WFC3 provides access to many important rest-frame optical emission lines over a wide range of redshifts, from $H\alpha$ down to $z=0.7$ to [OII] $\lambda3727$ at $z=3.4$.  Figure \ref{fig-redshift-range} shows the detectability of emission lines with 2-orbit depth G141 grism observations at various redshift ranges.
%(see Figure \ref{fig-redshift-range}).  
Despite the low spectral resolution ($R\sim130$) of the WFC3 grism, the resulting redshift accuracy is an order of magnitude better than typical photometric redshift errors \citep{2012ApJS..200...13B}. 

In this paper, we combine imaging and photometric redshifts from the Cosmic Assembly Near-Infrared Deep Extragalactic Legacy Survey \citep[CANDELS;][]{2011ApJS..197...35G,2011ApJS..197...36K} and WFC3/IR grism observations from the 3D-HST survey \citep{2012ApJS..200...13B} to produce a new grism spectroscopic redshift catalog for \emph{H}-band selected galaxies in the GOODS-South field.  The catalog contains emission-line redshifts for 608 sources which have no previously published spectroscopic redshifts and contains 234 new redshifts at $z>1.5$.  The paper is organized as follows: in Section 2 we introduce the datasets used in constructing the redshift catalog.  In Section 3 we present the methodology used to inspect the grism spectra and measure redshifts.  In Section 4 we present an overview of the redshift catalog and its key properties, provide an analysis of the accuracy of the redshift measurements, and use the new redshifts to identify close galaxy pair candidates.  Finally, in Section 5 we summarize our work. Throughout this paper, we adopt the Chabrier IMF and the following cosmology: $H_{0}=70$ $kms^{-1}Mpc^{-1}$, $\Omega_{M}=0.3$, $\Omega_{\Lambda}=0.7$. All magnitudes are in the AB system.

  \section{Observations and Sample Selection}
  
  \subsection{Optical and Infrared Imaging}
  
\begin{figure}
\epsscale{1.17}
\plotone{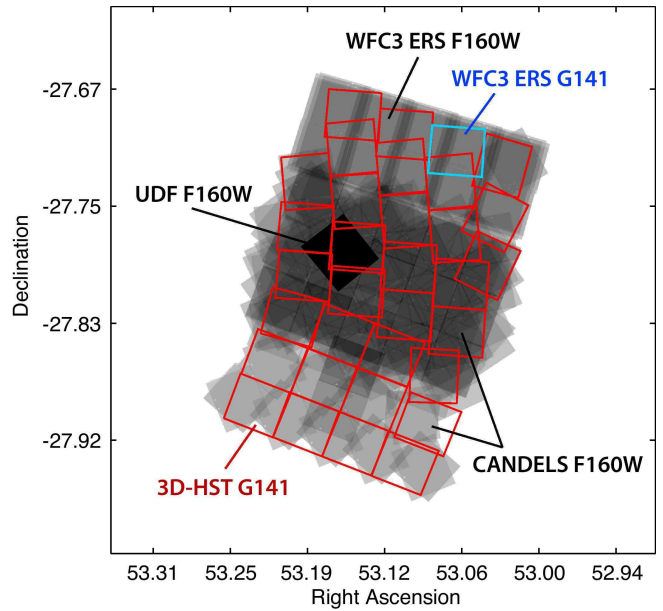}
\caption{Layout in the GOODS-S field of the WFC3 F160W imaging and G141 grism observations used in this study.  The imaging comes from the CANDELS, WFC3 ERS and HUDF09 programs, while the grism observations come from the 3D-HST (red) and WFC3 ERS programs (blue).
\label{fig-exp-map}}
\end{figure}

Our parent sample is drawn from the \emph{H}-band selected photometric catalog of \citet{2013ApJS..207...24G}, which made use of \emph{HST}/WFC3 imaging of the GOODS-S field from three programs: CANDELS, the WFC3 Early Release Science program \citep[ERS;][]{2011ApJS..193...27W}, and the HUDF09 program \citep{2010ApJ...709L.133B}. The location of the WFC3 imaging from these three programs is shown in Figure \ref{fig-exp-map}.  CANDELS has observed GOODS-S using a two-tiered Wide+Deep strategy.  The Deep region covers the central third of the GOODS-S area \citep[55 arcmin$^2$;][]{2004ApJ...600L..93G} with 3, 4, and 6 orbits of F105W, F125W, and F160W imaging, respectively.  The Wide region covers the southern third of the field with 2-orbit depth imaging in all three bands.  The ERS program covers the northern third of GOODS-S with 2-orbit depth imaging in the F098M, F125W, and F160W bands.  Finally, an area of 4.6 arcmin$^2$ in GOODS-S, the Hubble Ultra Deep Field, is covered by very deep 24, 34, and 53 orbits of F105W, F125W, and F160W imaging.  The CANDELS team carried out a consistent reduction of the WFC3 imaging from all of these programs; for details we refer readers to \citet{2011ApJS..197...36K}.
  
The GOODS-S field has also been observed in the optical with the Advanced Camera for Survey (ACS) on \emph{HST} as part of the GOODS \emph{Hubble} Treasury Program
(P.I.~M.~Giavalisco) in the $B, v, i,$ and $z$ bands with a total exposure time of 7200, 5450, 7028, and 18,232 seconds.  For this study, we made use of the publicly available, version v3.0 mosaicked images from the GOODS Treasury Program.  In the mid-infrared, we make use of \emph{Spitzer}/IRAC 3.6 and 4.5 $\mu$m imaging from the \emph{Spitzer} Extended Deep Survey \citep[SEDS; P.I.~G.~Fazio;][]{2013ApJ...769...80A}, which reaches total $3\sigma$ depths of $\sim26$ AB mag.

In addition to the observations described above, the GOODS-S region has been targeted for some of the deepest ground-based observations ever taken, ranging from the \emph{U} band \citep{2009ApJS..183..244N} to the \emph{K} band \citep{2014arXiv1409.7082F}. A detailed description of these datasets can be found in \citet{2011ApJS..197...35G} and \citet{2013ApJS..207...24G}.

\subsection{Photometry and Photometric Redshifts}
  
  For this study, we make use of the CANDELS photometric catalog of \citet{2013ApJS..207...24G}.  The catalog is \emph{H}-band selected using a ``max-depth" image that combines all available F160W in the GOODS-S field.  The catalog contains 34930 unique sources and is 50\% complete at $H\sim26$.  Multiwavelength photometry is obtained for the available \emph{HST} bands using a modified version of SExtractor (\citealt*{1996A&AS..117..393B}; see \citealt{2013ApJS..206...10G}) using the F160W observations as the detection image.  For \emph{Spitzer}/IRAC imaging and ground-based observations of mixed resolution, the TFIT software \citep{2007PASP..119.1325L} was employed to obtain PSF-matched photometry.  Further details on the CANDELS multiwavelength photometry catalogs can be found in \citet{2013ApJS..207...24G} and \citet{2013ApJS..206...10G}.

  Photometric redshifts for each source were generated from SED modeling using the photometry catalog of \citet{2013ApJS..207...24G}. A hierarchical Bayesian approach was employed in which the full photometric redshift probability distribution from 11 independent CANDELS investigators are combined to produce a more accurate redshift estimate.  The detailed description of this process can be found in \citet{2013ApJ...775...93D}. The photometry used ranged from the \emph{U}-band to the \emph{Spitzer}/IRAC 4.5 $\mu$m filter; see \citet{2013ApJS..207...24G}.  The resulting photometric redshifts are found to be accurate to the 2.9\% level and have an outlier fraction (OLF) of 9.1\% when compared with a sample of available spectroscopic redshifts.

  \subsection{WFC3/IR Grism Data}
  
 GOODS-S contains near complete spectroscopic coverage in the NIR with 2-orbit depth \emph{HST}/WFC3 G141 grism observations taken by the 3D-HST survey \citep[PI: P. van Dokkum;][]{2012ApJS..200...13B} and the WFC3 ERS program \citep{2011AJ....141...14S}, corresponding to a limiting line flux of $3\times10^{-17} {\rm erg/s/cm^{2}}$. The locations of the G141 observations in the GOODS-S field are shown in Figure \ref{fig-exp-map}.  The publicly available data were reduced using the aXe software package \citep{2009PASP..121...59K} to produce two- and one-dimensional wavelength- and flux-calibrated spectra. Spectra were reduced using the default (V2.0) aXe parameters.  This means we use a single sky background and do not account for the background fluctuations which typically affect the WFC3 grisms \citep{2014wfc..rept....3B}. (The GOODS-S G141 observations have lower overall background than the other CANDELS / 3D-HST fields, although the GOODS-S background can vary from 1--2 e$^-$/s within a single pointing, see Appendix B of Brammer et al.~2012.) The extraction window was set to be four times the object size projected perpendicular to the dispersion direction, where object size is measured from the F140W image using SExtractor. The spectra each cover a wavelength range of $1.1\leq \lambda \leq 1.7 \mu m$ with a resolving power $R\sim 130$ (46.5\AA{}/pixel) for point sources. For each observation with the grism, an accompanying direct F140W image is taken to determine the wavelength zero-point for each spectrum. The uncertainty in the zero-point and the dispersion are 8\AA{} and 0.06\AA{}/pixel respectively. The dispersion correlates to $\sim 1000 km s^{-1}$ for H$\alpha$ at $z> 1$. The total exposure time for each F140W direct image is 812 s and the total exposure time for each G141 grism image ranges between 4511 - 5111 s.  

Finally, we registered the grism observations to the CANDELS imaging in the field by running SExtractor on the F140W direct images and cross-matched the resulting source catalog to the CANDELS F160W catalog of \citet{2013ApJS..207...24G}.  Through this cross-matching we derived an astrometric correction for each individual 3D-HST tile; the average derived correction was 0\farcs{163} in $\Delta\alpha$ and 0\farcs{248} in $\Delta\delta$. 
  
  \subsection{Sample Selection}
  
Our initial sample consists of 4511 sources selected from the catalog of \citet{2013ApJS..207...24G} based on the following criteria:
  
  \vspace{0.1in}
  \hangindent=0.25in \hangafter=1 $\bullet$ \emph{H}-band Magnitude: $H_{\rm F160W}\leq 24$

  \hangindent=0.25in \hangafter=1 $\bullet$ Photometric Redshift: $z_{phot}\geq 0.6$
  \vspace{0.1in}
  
\noindent These criteria are chosen such that prominent emission features fall within the sensitivity window of the G141 grism, but also such that the number of sources to be inspected does not become unmanageably large(e.g. increasing the magnitude cut to $H_{\rm F160W}\leq 25$ more than doubles the sample size.) 
It is worth noting that as a result of these selection criteria, sources that are continuum-faint but have high-equivalent width emission lines may be missed from our initial selection.  The same is true for sources with catastrophic failures in their photometric redshift estimates.

To determine which sources in our initial sample have preexisting spectroscopic redshifts, we compared the sample against a recent compilation of published spectroscopic redshifts in the GOODS-S field (N. Hathi, private communication). This compilation contains redshifts from various sources including \citet{2010A&A...512A..12B,2012MNRAS.425.2116C,2001MNRAS.328..150C,2004ApJ...600L.127D,2009ApJ...700..183H,2007ApJ...669..776K,2012yCat..35499063K,2004A&A...428.1043L,2005A&A...437..883M,2007A&A...465.1099R,2010ApJS..191..124S,2004ApJ...613..200S,2004ApJS..155..271S,2013ApJ...763L...6T,2004ApJ...601L...5V,2008A&A...478...83V,2009ApJ...695.1163V,2004A&A...421..913W,2009ApJ...706..885W,2011ApJS..195...10X}, and the ESO GOODS/CDF-S Master Catalog\footnote{Available online at \url{http://www.eso.org/sci/activities/garching/projects/goods/MasterSpectroscopy.html}}.  We refer to this compilation as the Master Spectroscopic Catalog hereafter. Based on this comparison, we define two samples: a primary sample consisting of 3007 sources which do not appear in the master spectroscopic catalog and a secondary sample of 1504 sources which have published spectroscopic redshifts. In the following sections, we analyze the grism spectra of both samples in an identical manner (i.e.,~with no prior knowledge of any published spectroscopic redshift) and use the secondary sample to test the accuracy of our grism-derived redshifts (see \S4.2).

\begin{figure*}[h]
\vspace*{-6in}
\epsscale{1.45}
\plotone{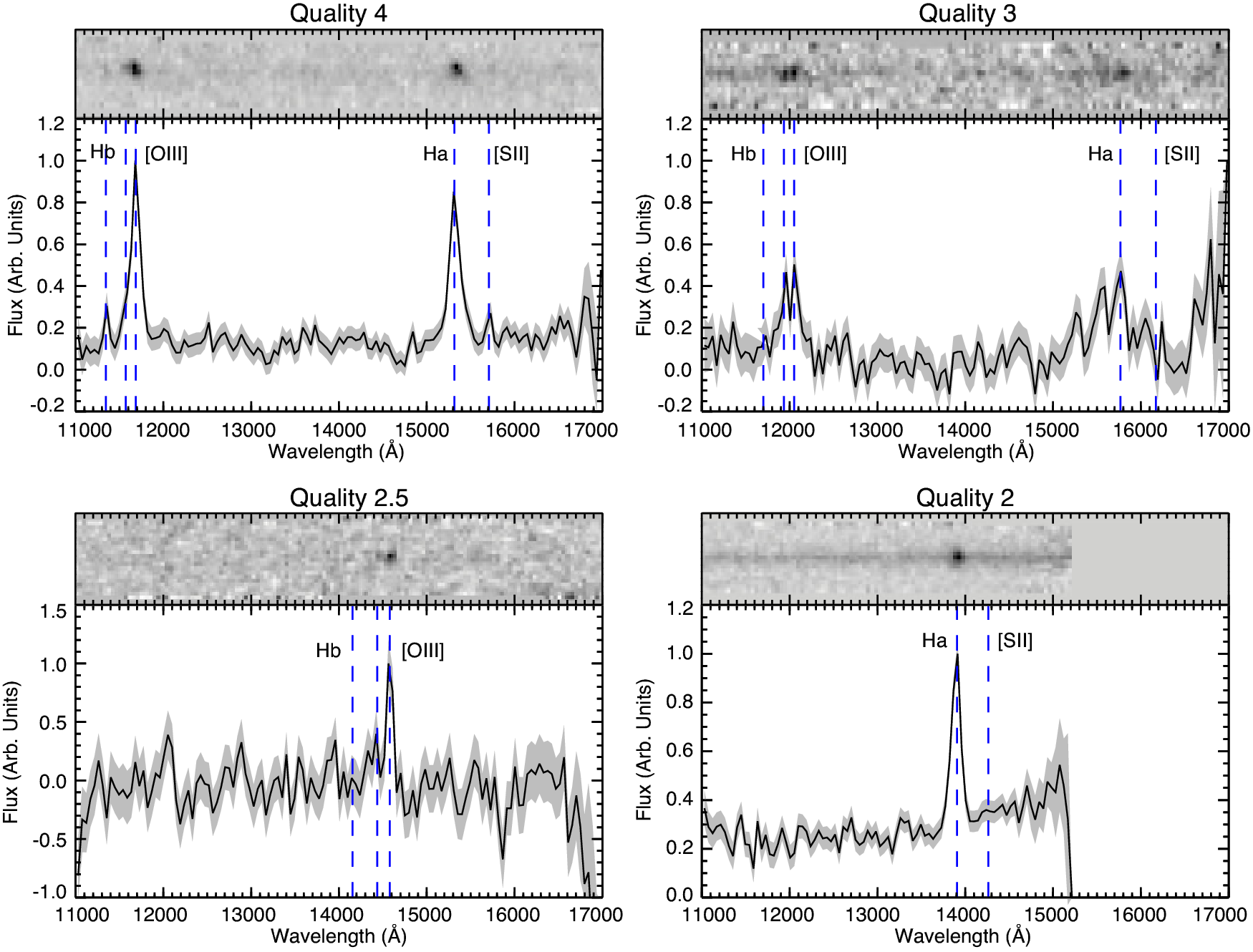}
\caption{Examples of G141 grism spectra with quality flags 4.0 (multiple high S/N emission lines), 3.0 (multiple, lower S/N emission lines), 2.5 (single emission feature; redshift agrees with photometric redshift), and 2.0 (single emission feature; redshift disagrees with photometric redshift). Vertical dashed lines indicate the location of major emission features.  \label{fig-conf4-ex}}
\end{figure*}
	    
Of the 4511 sources in the initial sample, 2314 sources in the primary sample and 1226 sources in the secondary sample fall within the 3D-HST G141 footprint and are detected in the F140W imaging.  For the sources in the primary sample, we extracted 2723 unique grism spectra from the 36 individual 3D-HST and ERS pointings, with 343 sources being identified in multiple pointings.  We extracted 1464 unique spectra for the secondary sample, with 224 sources being identified in multiple pointings.
  
\begin{figure}[t]
\epsscale{1.15}
\plotone{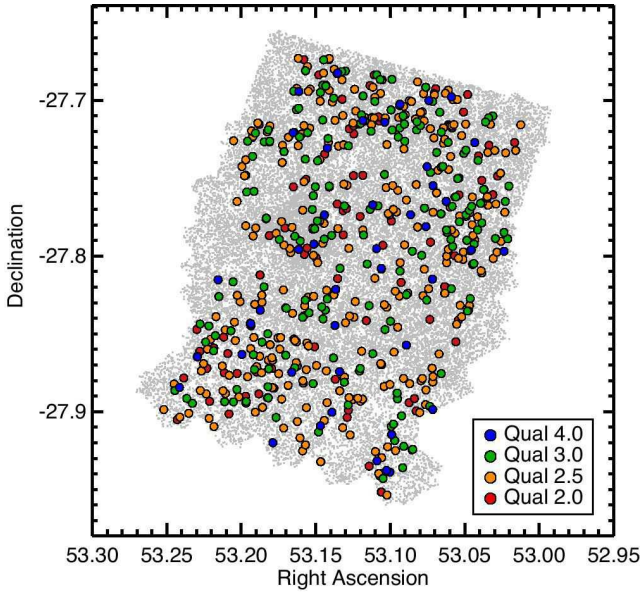}
\caption{Distribution of sources with new grism redshifts in GOODS-S.  Grey points are all sources from the catalog of Guo et al.~(2013) and filled circles are sources with grism redshifts, color coded according to their assigned quality; see \S3 for details.  \label{fig-source-qual}}
\end{figure}

\section{Redshift Measurement}
	
  The extracted 2D and 1D spectra for each object in our samples were visually inspected using the {\tt SpecPro} IDL software package \citealp{2011PASP..123..638M}. In addition to the spectrum of each source, {\tt SpecPro} also provides the user with the ability to display thumbnail images of the source being inspected, a plot of available photometry with an accompanying SED fit, and the location and width of the extraction window. To aid in the identification of emission lines, {\tt SpecPro} provides the predicted location of various emission and absorption features based on the input photometric redshift for the source.  Source spectra were also inspected for contamination from overlapping spectra using an estimate supplied by aXe which is indicated atop the 1D spectrum. It should be noted that no attempt was made to model or subtract contamination from each spectrum, although emission lines from neighboring sources were identified using the photometric or spectroscopic redshift of the contaminating source and contamination from zeroth order spectra is indicated by the contamination estimate supplied by aXe.  Any source which exhibited excessive contamination or had data quality issues (e.g.~significantly incomplete spectrum due to the spectrum being dispersed off the edge of the detector or spectra which overlap with defective portions of the detector) were removed from the sample.  These sources accounted for roughly 17\% of all 4187 extracted spectra in the primary and secondary samples.
  
  During inspection, any visible spectral features were roughly fit manually and subsequently fit via cross-correlation with spectral templates provided in the software to determine the redshift of the source. {\tt SpecPro's} cross-correlation method is adapted from the cross-correlation routines originally written for the SDSS spectral reduction package, which follows the technique of \citealt{1979AJ.....84.1511T}. When the automated cross-correlation failed, the redshift was determined manually by fitting the peaks of the template emission features to those observed in the grism spectrum. This was done for less than 5\% of sources. The templates used for cross-correlation are taken from the VVDS \citep{2005A&A...439..877L} and PEGASE \citealp{1997A&A...326..950F} templates. The emission lines most often used for identification were H$\alpha$/[NII]$\lambda\lambda$6550+6584 and [SII]$\lambda\lambda$6717+6731, H$\beta$ and [OIII]$\lambda\lambda$4959+5007, and [OII]$\lambda$3727. The spectral resolution of the grism is such that the above sets and duplexes are not resolved, but it is enough to produce certain distinguishing profiles that aid in differentiating each from other strong lines (e.g.~the asymmetric profile of [OIII]$\lambda\lambda$4959+5007.)
	
   Upon inspection, the derived redshift of each source was assigned a quality flag based on the strength and number of the identified emission lines and the agreement with existing photometric redshift estimates.  If a source was identified in multiple pointings and therefore assigned multiple redshifts, the higher quality redshift was always chosen. In the case where multiple redshifts of the same quality exist, the redshift of the source was taken to be the average of the individual redshifts. The quality scheme for the derived redshifts is as follows:
   
  \vspace{0.1in}
  \hangindent=0.25in \hangafter=1 $\bullet$ 4.0: Multiple high S/N emission lines 
  
  \hangindent=0.25in \hangafter=1 $\bullet$ 3.0: Combination of high and low S/N emission lines. 
  
  \hangindent=0.25in \hangafter=1 $\bullet$ 2.5: Single high S/N emission line and redshift agrees with 68\% confidence interval of photometric redshift 
  
  \hangindent=0.25in \hangafter=1 $\bullet$ 2.0: Single high S/N emission line and redshift does not agree with 68\% confidence interval of photometric redshift
  
   \vspace{0.1in}
   
\noindent Examples of spectra correlating to each quality flag can be seen in Figure \ref{fig-conf4-ex}.  While quality 3.0 and 4.0 redshifts are the most reliable, given the multiple emission lines identified, we show in \S4.2 that sources assigned a quality of 2.0 or 2.5 demonstrate excellent agreement with prior spectroscopic redshift measurements.

\section{Results}
   
\subsection{Catalog Properties}
      
   Upon inspection and classification of the 2411 grism spectra in the primary sample, we have identified 608 sources with visible emission lines for which redshifts could be measured. Of these, 45 exhibited multiple high S/N emission lines (quality 4.0), 181 exhibited multiple emission lines with some low S/N emission lines (quality 3.0), 293 exhibited a single high S/N emission line whose redshift agrees with the 68\% confidence interval of the CANDELS photometric redshift estimate (quality 2.5), and 89 exhibited a single high S/N emission line which does not agree with the 68\% confidence interval of the photometric redshift (quality 2.0). In the secondary sample, we have identified 411 sources with visible emission lines for which redshifts could be measured. Of these, 35 exhibited multiple high S/N emission lines (quality 4.0), 167 exhibited multiple emission lines with some low S/N emission lines (quality 3.0), 157 exhibited a single high S/N emission line whose redshift agrees with the 68\% confidence interval of the CANDELS photometric redshift estimate (quality 2.5), and 52 exhibited a single high S/N emission line which does not agree with the 68\% confidence interval of the photometric redshift (quality 2.0).

\begin{figure}[t]
\epsscale{1.15}
\plotone{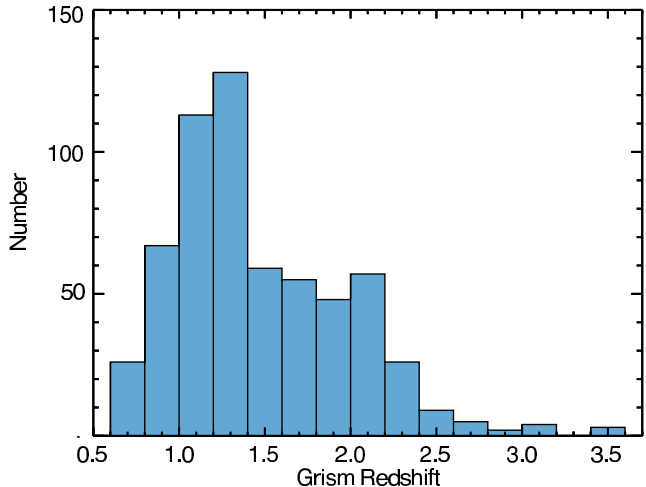}
\caption{Distribution of the 608 new grism redshifts in GOODS-S. The redshfits span a range of $0.677\leq z\leq 3.456$ and have a median redshift of $z=1.282$. There are 234 sources with $z\ge1.5$ \label{fig-grism-zdist}} 
\end{figure}
 
\begin{figure}[t]
\epsscale{1.16}
\plotone{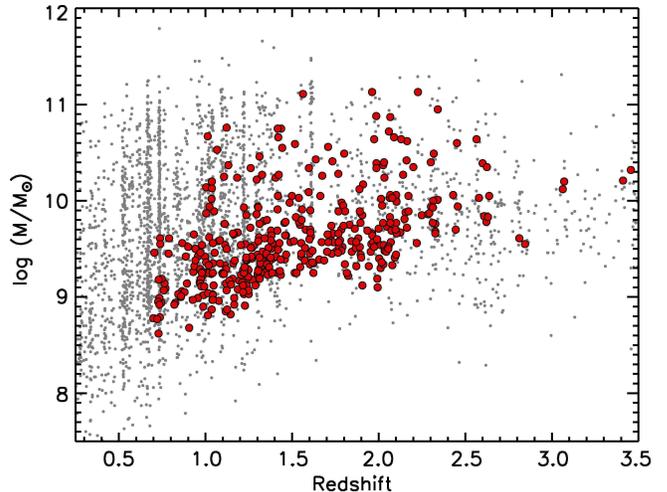}
\caption{Mass distribution as a function of spectroscopic redshift for galaxies in the GOODS-S field.  Galaxies with new grism redshifts from this work are shown as red circles, while galaxies with previously published spectroscopic redshifts are shown as grey points. \label{fig-mass-redshift}}
\end{figure}

The final catalog contains 1019 grism redshifts for galaxies brighter than $H=24$ in the GOODS-S field.  Roughly 60\% (608/1019) of the redshifts are new, in that these galaxies have no previously published spectroscopic redshift.  The new redshifts span a range of $0.677\leq z\leq 3.456$ and have a median redshift of $z=1.282$.  The catalog contains a total of 234 new redshifts for galaxies at $z\ge1.5$.  The spatial distribution of the 608 galaxies with new redshifts can be seen in Figure \ref{fig-source-qual} and their redshift distribution is shown in Figure \ref{fig-grism-zdist}.  In addition, the stellar mass distribution of these 608 galaxies is shown in Figure \ref{fig-mass-redshift}.  Here masses are calculated by SED modeling using photometry from the \citet{2013ApJS..207...24G} catalog as described in Mobasher et al.~(2014, in prep.).  We find that galaxies with new grism redshifts in our catalog are, on average, three times less massive than their counterparts with literature redshifts at $z\sim1-2$.   

\begin{figure*}[t]
\epsscale{1.1}
\plotone{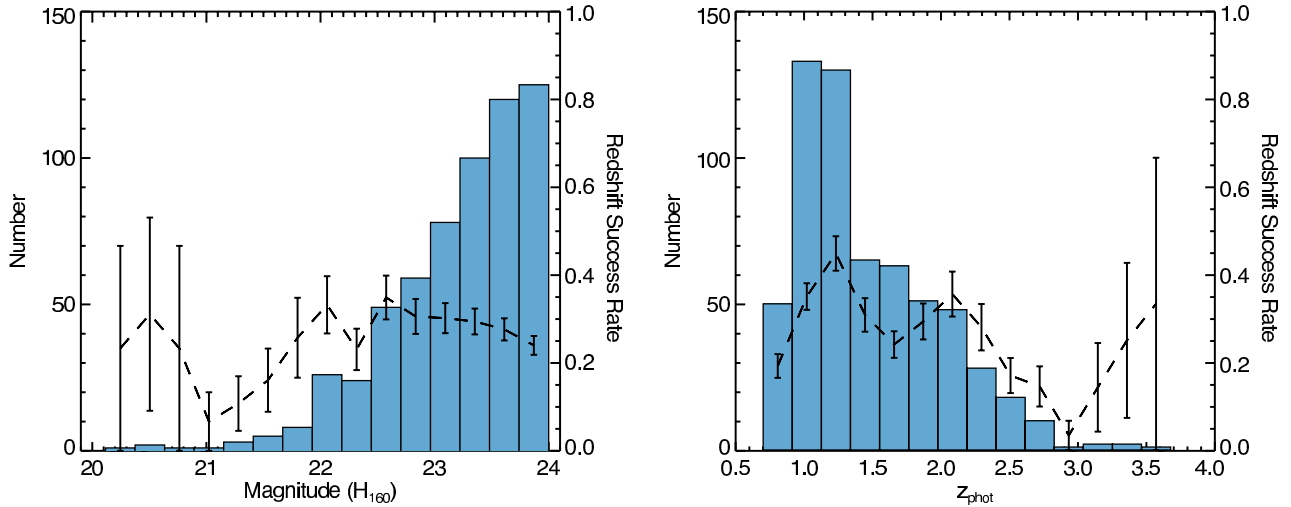}
\caption{\emph{(left)} Magnitude distribution of sources with new grism redshifts. The dashed line represents the redshift ``success'' rate in each bin with error bars given by number statistics based on the number of sources in that bin. \emph{(right)} Redshift distribution of sources with new grism redshifts. \label{fig-success-rate}}
\end{figure*}

Over the magnitude range of our primary sample ($H<24$), we find that our ability to successfully measure a redshift is not strongly dependent on the $H$-band magnitude of the source.  Figure \ref{fig-success-rate} shows the magnitude distribution of sources in our primary sample along with the redshift success rate in each magnitude bin (defined as the ratio of the number of sources with grism redshifts of $\rm{quality}\geq2.0$ to the number of all sources in our initial sample in a given bin).  Over the magnitude range $22<H<24$, our success rate ranges from 20 to 30\%, showing only a mild decrease for our faintest sources.  Also shown in Figure \ref{fig-success-rate} is our success rate as a function of redshift.  Here we find a steady decrease from 35\% to 10\% in the redshift range $2.0<z<2.75$.  This is likely due to [OIII] and H$\beta$ shifting beyond $1.7\mu m$ at $z=2.4$ and 2.5, respectively, leaving [OII] as the single emission line visible in the G141 sensitivity window at $z>2.5$.

\begin{figure}[t]
\epsscale{1.15}
\plotone{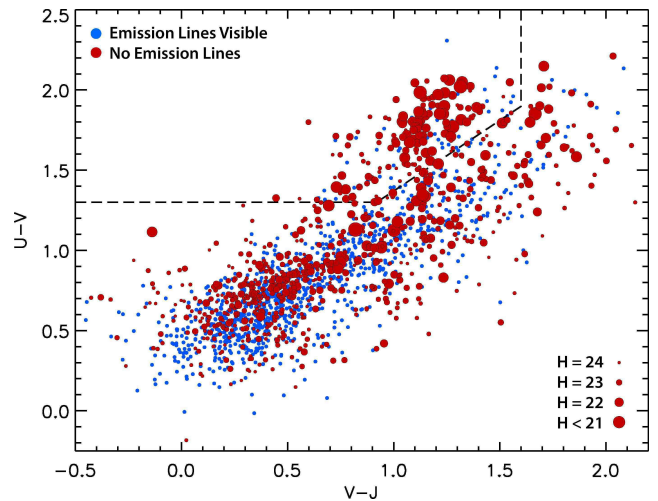}
\caption{UVJ plot of rest-frame colors for all sources in both the primary and secondary samples. Blue points represent sources for which a grism redshift was successfully obtained and red sources represent sources for which a grism redshift was not obtained with point size scaled according to the source's \emph{H}-band magnitude. The dashed line separates the star-forming "blue cloud" and the quiescent "red sequence". We find that sources without grism redshifts tend to be bright quiescent systems or faint star forming galaxies. \label{fig-uvj}}
\end{figure}

We have investigated the nature of the sources for which a grism redshift could not be determined due to the lack of visible emission lines and find them to be a combination of bright, quiescent galaxies and faint star-forming systems.  In Figure \ref{fig-uvj}, we show the location of these galaxies on a UVJ diagram \citep{2009ApJ...691.1879W}.  This phase-space separates blue, star forming galaxies from redder systems that are heavily dust extinguished or passively evolving.  For this analysis, rest-frame colors were determined with the EAZY code \citep{2008ApJ...686.1503B} using the observed photometry from Guo et al.~(2013) and the CANDELS photometric redshift catalog.  Of the galaxies which lack visible emission lines, we find that 33.0\% are dusty or quiescent ($U-V_{\rm rest}>1.3$).  This is nearly three times greater than the 12.4\% of galaxies with measured grism redshifts that have similar rest-frame colors.  The remaining 67.0\% are blue, star-forming systems ($U-V_{\rm rest}<1.3$) that are predominately faint (64.5\% are fainter than $H\sim23$; the same is true for only 23.7\% of the passive/dusty population).  We therefore conclude that objects which failed to yield a grism redshift are largely a combination of quiescent galaxies that lack emission lines and star-forming galaxies with emission lines fainter than the detection limit of the grism observations.
    
\subsection{Redshift Accuracy}
    
\begin{figure*}
\epsscale{1.0}
\plotone{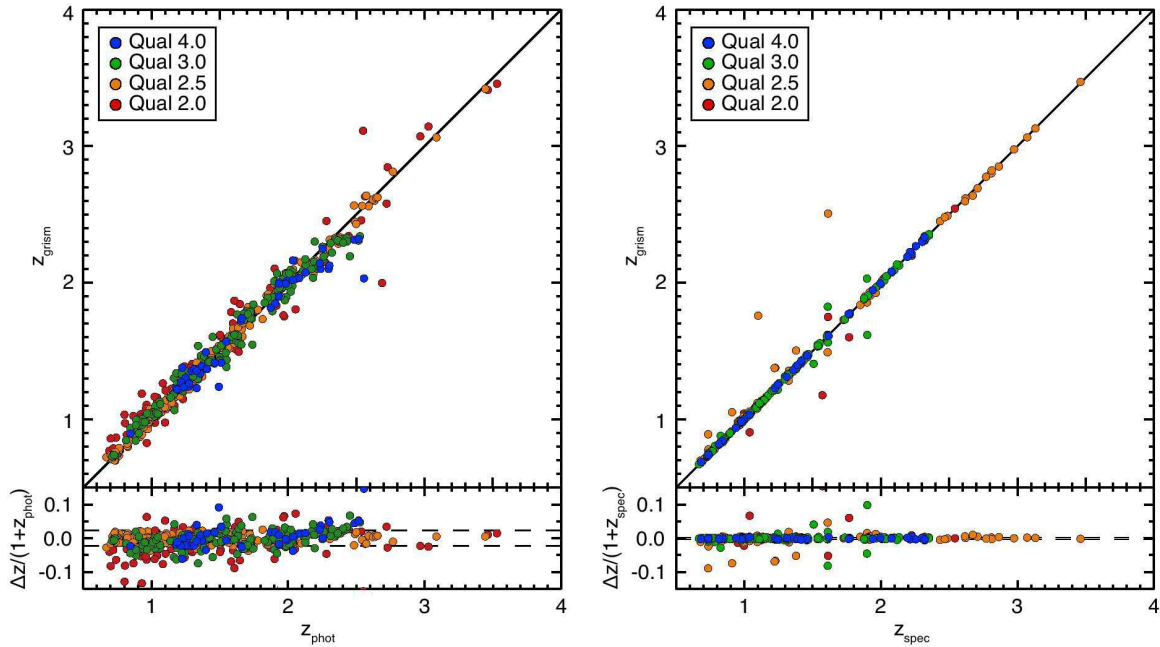}
\caption{\emph{(left)} Grism redshifts from this work versus CANDELS photometric redshifts. The solid line represents $z_{grism} = z_{phot}$, while the dashed lines represent $\sigma_{\rm NMAD}=0.0236$. \emph{(right)} Grism redshifts from this work versus ground-based spectroscopic redshifts, with $\sigma_{\rm NMAD}=0.0028$.  \label{fig-z-comp}}
\vspace{0.2in}
\end{figure*}

    To evaluate the accuracy of our measured redshifts, we first examine the agreement of the new grism redshifts with the CANDELS photometric redshifts.  This is shown in the left panel of Figure \ref{fig-z-comp}. We to quantify the accuracy of the catalog via the parameter $\sigma_{NMAD}$, defined as $1.48\times median(|\Delta z|/(1+z_{phot}))$. The grism redshifts show good agreement with the photometric redshifts, with $\sigma_{NMAD}=0.0236$ and an OLF ($|\Delta z|/(1+z_{phot})\ge0.15)$ of 0.0150. These values are comparable to the accuracy of the photometric redshifts when compared against ground-based spectroscopic redshifts, so we may reasonably assume that the majority of this scatter is due to the error in the photometric redshifts. We believe many of the outliers in this evaluation are sources with SEDs contaminated by strong emission lines. A comparison to the work of \citet{2014arXiv1409.7119H}, which takes into account intermediate-band $\emph{Subaru}$ photometry \citep{2010ApJS..189..270C} and the contribution of emission lines in addition to the photometry presented in \citet{2013ApJS..207...24G}, reduces the number of outliers in the sample from 9 to just 2.

  To further assess the accuracy of the grism redshifts, we examine the agreement of the secondary sample with the redshifts conained within the master spectroscopic catalog. Each extracted spectrum was inspected using the same methods employed for the primary sample; i.e.,~we used the photometric redshifts to aid in our redshift determination but had no knowledge of the published spectroscopic redshifts.  This process resulted in 411 successful grism redshift determinations.  A comparison of the grism and ground-based spectroscopic redshifts is shown on the right panel of Figure \ref{fig-z-comp}.  This comparison yields a scatter of $\sigma_{NMAD}=0.0028$ with an OLF of 0.0098.  Compared to the work of \citet{2012ApJS..200...13B}, we find excellent agreement with the accuracy reported by the 3D-HST team ($\sigma_{NMAD}=0.0035$). 

  As a further test of the accuracy of our catalog, we have compared our grism redshifts to those obtained by the WFC3 ERS program. Straughn et al. obtained 48 redshifts via their G102 and G141 observations, of which ten sources with the highest quality redshifts meet our magnitude and redshift selection criteria and received quality flags of 2.0 or higher in our inspection. We see excellent agreement between our results and those of Straughn et al., with a comparison in the manner described above giving a result of $\sigma_{NMAD}=0.0016$. 

  We find that the scatter between the grism and ground-based spectroscopic redshifts is not significantly increased ($\Delta\sigma_{NMAD}\approx0.0001$) for those sources whose redshifts were fit manually versus those fit via the built-in cross-correlation routines in {\tt SpecPro}.  We also find that the scatter does not vary significantly with quality flag, ranging from $\sigma_{NMAD}=0.0026$ for quality 4 redshifts to $\sigma_{NMAD}=0.0033$ for quality 2 redshifts.  We therefore propose quality 2.0 as the minimum reliable quality for this redshift catalog. In addition, we find no correlation between scatter and the effective radius of the sources as defined in \citet{2014ApJ...788...28V}.

\subsection{Redshift Catalog}
 
Starting with a sample of 4511 sources, we have obtained a total of 1019 grism redshifts for galaxies brighter than $H=24$ in the GOODS-S field.  Of these, 608 are new redshift measurements for galaxies in our primary sample, which do not have previously published spectroscopic redshifts.  The coordinates and redshifts of all 1019 galaxies are listed in Table \ref{tab-redshifts}.  The details of the table columns are given below.

\vspace{0.1in}  
\indent    1. Source ID from Guo et al.~(2013)\\*
\indent    2. Right ascension (J2000)\\*
\indent    3. Declination (J2000)\\*
\indent    4. $H$-band magnitude (AB) from Guo et al.~(2013)\\*
\indent    5. Redshift derived from G141 grism spectrum\\*
\indent    6. Redshift from Master Spectroscopic Catalog\\*
\indent    7. Emission line(s) used for redshift determination\\*
\indent    8. Redshift quality flag (see \S3 for details)

\subsection{Newly Identified Galaxy Pair Candidates}
    
In this section, we highlight one of the potential uses for our redshift catalog, namely the identification of close galaxy pairs at $z>1$.  Due to the slitless nature of the WFC3 grism, we can now dramatically increase the spectroscopic sampling of star forming galaxies at these redshifts.  Even with a spectral resolution of $\delta v \sim 1000 km s^{-1}$, the redshift accuracy of the G141 grism spectra ($\sigma_{\rm z}/(1+z)$ = 0.3\%) is far better than typical photometric redshift errors (3\%) and the accuracy of low-resolution prism spectroscopy (1.2\%), which have been used in the past to study the merger rate and environments of galaxies at $z\sim1$ and beyond \citealp[i.e.,][]{2011ApJ...735...53P,2012ApJ...746..162N}.

Combining the grism redshifts in our catalog with our master spectroscopic catalog, we have identified 20 galaxy pair candidates in GOODS-S with at least one member whose redshift comes from the new WFC3/G141 spectra.  To identify galaxy pairs, we inspect the neighbors of each galaxy with a grism redshift of quality 2.0 or greater and define close companions as those that are 1) within a projected distance of 50 kpc, and 2) have a redshift difference of $\Delta z/(1+z)<0.03$, or roughly $\delta v \sim 1000 km s^{-1}$ at $z>1$.  Based on these criteria, we find 20 galaxies with potential companions in the CANDELS/GOODS-S region.  Five of the pairs are comprised of two galaxies with new grism redshifts and 15 are grism sources with companions that appear in our master spectroscopic catalog. The sample spans a redshift range of $0.787\leq z\leq 2.33$, with four of the pairs identified at $z\sim2$.  This sample represents roughly a factor of two increase in the number of such pairs identified with the master spectroscopic catalog alone. On average, the new companions to sources with spectroscopic redshifts are nearly one magnitude fainter in the \emph{H}-band, which highlights the ability of the grism to detect fainter objects than are usually seen via ground-based spectroscopy. In addition, the objects found in these pairs are approximately ten times less massive than objects typically observed in pair studies at this redshift \citep{2013A&A...553A..78L,2014A&A...565A..10T} The coordinates and redshifts of the newly detected pair candidates are listed in Table \ref{tab-pairs}. 
    
\section{Summary}
  
We have constructed a redshift catalog for galaxies in the CANDELS/GOODS-S field using \emph{HST}/WFC3 G141 grism observations from the 3D-HST survey and WFC3 ERS program.  The G141 spectra cover a wavelength range of $1.1\leq \lambda \leq 1.7 \mu m$, which allows for the detection of prominent emission lines over a wide redshift range, from $H\alpha$ at $z=0.7$ to [OII] $\lambda3727$ at $z=3.4$.  Our catalog is \emph{H}-band selected based on the CANDELS photometry catalog of \citet{2013ApJS..207...24G}.   Spectra were extracted for all GOODS-S sources which are brighter than $H=24$ and have a photometric redshift $z_{phot}\geq 0.6$.  Each spectrum was visually inspected, emission lines were identified with the aid of CANDELS photometric redshifts, and redshifts were measured via cross-correlation with empirical spectral templates.  Derived redshifts were assigned a quality ranging from 4.0 for sources with multiple strong emission lines, to 2.0 for sources with a single visible emission line.  The resulting catalog contains new grism redshifts for 608 galaxies which have no previously published spectroscopic redshift in the GOODS-S field.  These redshifts span a range of $0.677\leq z\leq 3.456$ and include 234 new redshifts for galaxies at $z\ge1.5$.  The catalog also contains grism redshifts for 411 galaxies which have existing redshifts in the literature.
 
We find good agreement between our grism-derived redshifts and existing photometric redshifts from CANDELS ($\sigma_{NMAD}=0.0236$).  We've also tested the accuracy of our redshifts by extracting and inspecting the spectra of GOODS-S sources with published spectroscopic redshifts.  This analysis was done blind, with only the photometric redshift of each source known during the inspection.  Here we find excellent agreement between our redshifts and the published values ($\sigma_{NMAD}=0.0028$).  This agreement holds even for redshifts measured with only a single emission line (quality 2.5 and 2.0 in the catalog). 

Finally, we use our redshift catalog to identify 20 new galaxy pair candidates at $z=1-2$.  These were chosen to have a projected separation of 50 kpc and a velocity offset of $\delta v\sim1000 km s^{-1}$.  Included in this sample are four new pairs identified at $z\sim2$.

\vspace{0.25in}
This work is based on observations taken by the CANDELS Multi-Cycle Treasury Program (GO 12177) and the 3D-HST Treasury Program (GO 12328) with the NASA/ESA HST, which is operated by the Association of Universities for Research in Astronomy, Inc., under NASA contract NAS5-26555 and by the Spitzer Space Telescope, which is operated by the Jet Propulsion Laboratory, California Institute of Technology under a contract with NASA. We also acknowledge partial support from NSF 0808133 and HST-AR 12822.03-A.

\clearpage

\LongTables
\begin{deluxetable*}{cccccclc}
%\begin{deluxetable}{cccccclc}
\tabletypesize{\tiny}
\tablecaption{Grism Redshift Catalog of Sources in GOODS-S \label{tab-redshifts}}
\tablecolumns{8}
\tablewidth{0pt}
\tablehead{\colhead{Source ID} & \colhead{RA} & \colhead{Dec} & \colhead{AB(F160W)} & \colhead{$z_{grism}$} & \colhead{$z_{spec}$} & \colhead{Lines} & \colhead{Qual} \\ \colhead{(1)} & \colhead{(2)} & \colhead{(3)} & \colhead{(4)} & \colhead{(5)} & \colhead{(6)} & \colhead{(7)} & \colhead{(8)}}
\startdata
   36 &  03 32 24.52 & -27 57 12.8 & 23.54 & 0.733 & -- & H$\alpha$ & 2.5 \\
   54 &  03 32 25.47 & -27 57 05.7 & 23.95 & 1.805 & -- & [OIII] & 2.0 \\
   55 &  03 32 25.42 & -27 57 05.9 & 23.74 & 1.753 & -- & [OIII] & 2.0 \\
  173 &  03 32 25.39 & -27 56 33.9 & 22.97 & 0.683 & 0.684 & H$\alpha$, [SIII]$\lambda$9532 & 3.0 \\
  180 &  03 32 25.15 & -27 56 34.3 & 23.53 & 1.332 & -- & H$\beta$, [OIII], H$\alpha$ & 3.0 \\
  218 &  03 32 25.66 & -27 56 28.5 & 22.87 & 1.112 & -- & H$\alpha$ & 2.0 \\
  244 &  03 32 25.18 & -27 56 23.1 & 23.20 & 1.103 & -- & H$\alpha$ & 2.5 \\
  267 &  03 32 25.84 & -27 56 23.2 & 22.67 & 1.087 & -- & H$\alpha$ & 2.5 \\
\enddata
\tablecomments{[OII] and [OIII] in column 7 above refer always to the [OII]$\lambda\lambda$3727+3729 and [OIII]$\lambda\lambda$4959+5007 emission line doublets.}
\end{deluxetable*}

\clearpage

\begin{deluxetable*}{ccccccccccc}
\tabletypesize{\tiny}
\tablecaption{Grism-Identified Pairs in GOODS-S \label{tab-pairs}}
\tablecolumns{11}
\tablewidth{0pt}
\tablehead{\colhead{Source ID 1} & \colhead{Source ID 2} & \colhead{RA} & \colhead{Dec} & \colhead{RA} &\colhead{Dec} & \colhead{$z_{1}$} & \colhead{$z_{2}$} & \colhead{Angular Sep.} & \colhead{Proj. Sep.} & \colhead{Mass Ratio} \\ \colhead{(Guo+~2013)} & \colhead{(Guo+~2013)} & \colhead{Source 1} & \colhead{Source 1} & \colhead{Source 2} & \colhead{Source 2} & \colhead{} & \colhead{} & \colhead{(arcsec)} & \colhead{(kpc)} & \colhead{$M_{1}/M_{2}$} }
\startdata
  722 &   726 &  53.096496 & -27.925050 &  53.096385 & -27.925972 & 1.145 & 1.146 &  3.338 & 27.48 & 0.847 \\
 1708 &  1695 &  53.241265 & -27.903526 &  53.241173 & -27.903265 & 1.050 & 1.045 &  0.984 & 7.963 & 1.086 \\
 2524 &  2526 &  53.182411 & -27.890996 &  53.181927 & -27.891138 & 1.458 & 1.465 &  1.623 & 13.72 & 0.974 \\
 3606 &  3689 &  53.078473 & -27.878643 &  53.078256 & -27.878455 & 0.987 & 0.987 &  0.967 & 7.720 & 0.981 \\
 3707 &  3952 &  53.071376 & -27.877464 &  53.072763 & -27.876321 & 1.099 & 1.100 &  6.034 & 49.31 & 0.845 \\
 5948 &  5983 &  53.055988 & -27.855096 &  53.056169 & -27.855367 & 1.227 & 1.226 &  1.133 & 9.426 & 0.948 \\
 6223 &  6195 &  53.173546 & -27.852423 &  53.175354 & -27.852610 & 0.967 & 0.962 &  5.794 & 45.99 & 0.971 \\
 7129 &  7172 &  53.193578 & -27.843512 &  53.193960 & -27.843195 & 2.003 & 2.011 &  1.668 & 13.96 & 1.011 \\
\enddata
\end{deluxetable*}

\clearpage

\bibliography{ms.astroph.bib}
\end{document}